\documentclass[conference]{IEEEtran}

\usepackage[pdftex]{graphicx}

\usepackage{algorithm,algorithmic}
\usepackage{amsmath, amsfonts, amsthm,   url,array}
\graphicspath{../graphics}
\DeclareGraphicsExtensions{.pdf,.jpeg,.png}
\usepackage{booktabs, amssymb}
\usepackage{eqparbox}
\usepackage{mathtools}

\newcommand\Mycomb[2][n]{\prescript{#1\mkern-0.5mu}{}C_{#2}}

\usepackage{caption}
\usepackage{subcaption}

\usepackage[caption=false,font=footnotesize]{subfig}

\theoremstyle{definition}
\newtheorem{defn}{Definition}[section]

\theoremstyle{proposition}
\newtheorem{proposition}{Proposition}[section]

\begin{document}

\title{On Local and Global Centrality in Large Scale Networks}

\author{\IEEEauthorblockN{Sima Das}
\IEEEauthorblockA{Department of Computer Science\\
Missouri University of Science and Technology\\
Rolla, Missouri 65409.\\
sdp4b@mst.edu.}
}

\maketitle

 \begin{abstract}

Estimating influential nodes in large scale networks including but not limited to social networks, biological networks, communication networks, emerging smart grids etc. is a topic of fundamental interest. To understand influences of nodes in a network, a classical metric is {\em centrality} within which there are multiple specific instances including {\em degree centrality}, {\em closeness centrality}, {\em betweenness centrality} and more. As of today, existing algorithms to identify nodes with high centrality measures operate upon the entire (or rather global) network, resulting in high computational complexity. In this paper, we design efficient algorithms for determining the {\em betweenness centrality} in large scale networks by taking advantage of the modular topology exhibited by most of these large scale networks. Very briefly, modular topologies are those wherein the entire network appears partitioned into distinct modules (or clusters or communities), wherein nodes within the module (that likely share highly similar profiles) have dense connections between them, while connections across modules are relatively sparse. Using a novel adaptation of Dijkstra's shortest path algorithm, and executing it over local modules and over sparse edges between modules, we design algorithms that can correctly compute the {\em betweenness centrality} much faster than existing algorithms. To the best of our knowledge, ours is the first work that leverage modular topologies of large scale networks to address the centrality problem, though here we mostly limit our discussions to social networks. We also provide more insights on centrality in general, and also how our algorithms can be used to determine other centrality measures.

\end{abstract}

\IEEEpeerreviewmaketitle
\section{Introduction}

\label{sec:intro}

We are living in a world with massive scale social connections. It is not an exaggeration to say that in the last decade or so, the impact of online social communities to society is nothing short of radical. In order to better understand and reason about these networks, a significant body of work has emerged in the domain of modeling network evolution, identifying influences, detecting privacy breaches, improving network security and resilience and so on \cite{michalski:2011, viswanath:2011, olsson:2004, gross:2005, beach:2009}. Of fundamental importance to the analysis of large scale social networks is the measure {\em centrality} of a node (or vertex), which in very general terms measures its relative importance within the network (or graph) \cite{bavelas:1948, leavitt:1951}. In this paper, we are particularly interested in {\em betweenness centrality}, which in essence quantifies the number of times a node acts as a bridge along shortest paths between other nodes in the network. Finding the {\em betweenness centrality} is critical because several applications like, energy management in smart grids, finding terrorist cells through social networks, protein interaction and disease propagation in biological networks are all realizable depending on finding {\em betweenness centrality} nodes in the network. Unfortunately though, in large scale social networks finding the {\em betweenness centrality} is computationally expensive. In this paper, we design algorithms to reduce the associated computational cost of finding the {\em betweenness centrality} in large scale social networks by leveraging from the modular structure exhibited by several real world social networks.

\begin{figure}[!htb]
    \centering
  \hspace{4mm}  \includegraphics[scale=.4]{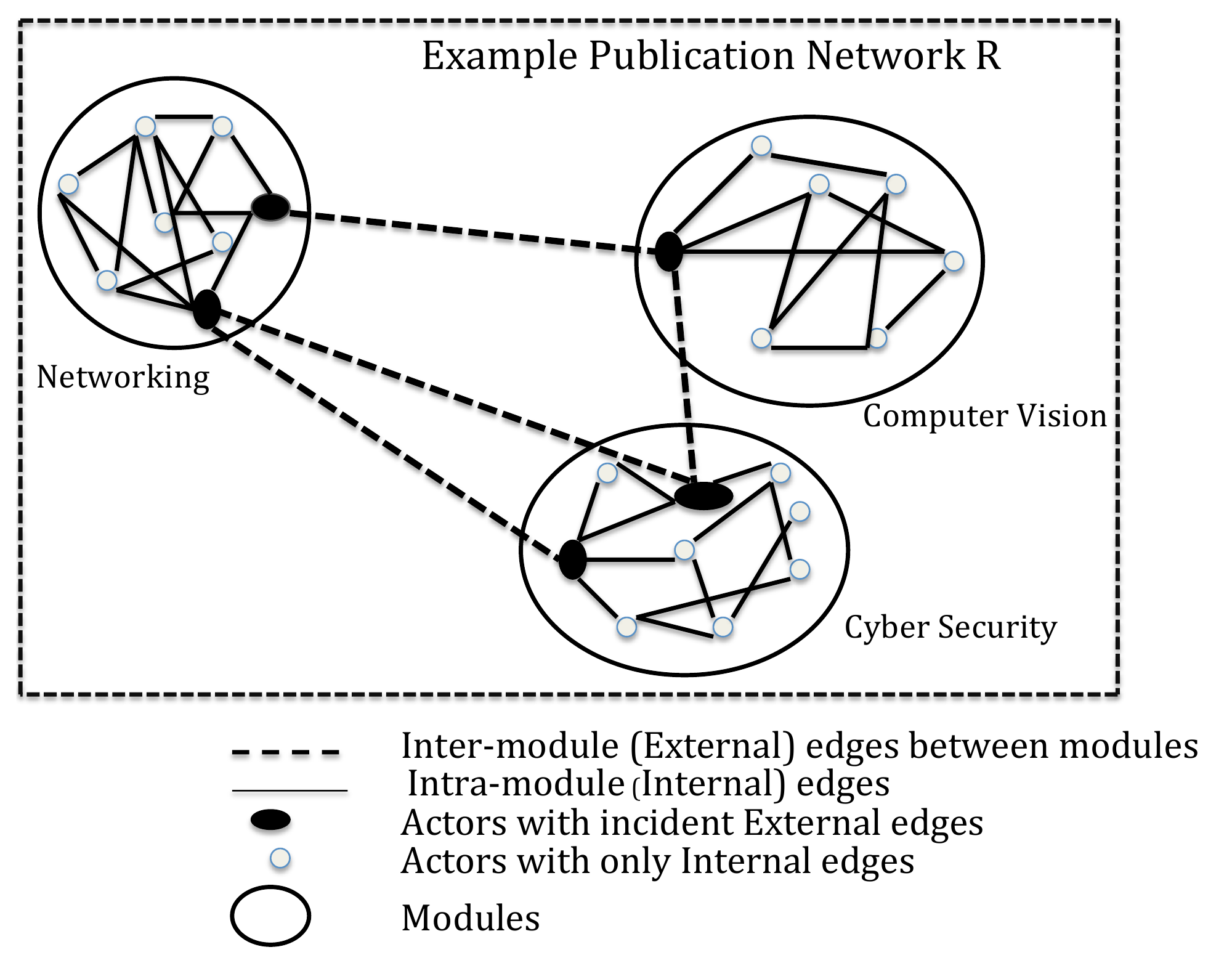}
    \caption{Modular structure over publication collaboration network $R$}
    \label{fig:modular}
\end{figure}

Topologies of complex networks in real world (including social networks) exhibit modular properties \cite{watts1999networks, hanneman2005introduction, mishra2007clustering, gupta2011evolutionary, bilgin2006dynamic}. For instance, it is easy to see that in a network like Facebook or Twitter, users will have multiple profiles and interests. Users sharing similar interests are likely to have dense interconnections amongst themselves (called a module), and naturally any user $U$ will be a part of multiple such communities. It is also natural to infer that there will be connections between users across communities, but such connections are much sparser since the number of users sharing all interests similar to user $U$ are likely a small number. In another such instance of modular network we consider collaboration in academic publication network. Consider three diverse areas of Computer Science - Computer Vision, Networking, Cyber Security. It is easy to infer that connections among authors specializing in each community are dense, while connections among authors across communities will be relatively much sparser, which happens under special instances of cross-disciplinary collaborations. A simple illustration of this phenomena is shown in Figure \ref{fig:modular}, where \emph{Internal edges} denote dense connections within a community, while \emph{External edges} denote sparser edges across communities. Needless to say analyzing {\em centrality} in such kinds of social networks is a problem of critical interests from the perspective of influence, load, information dissemination, resilience etc.

In this paper, we are focusing on {\em centrality} in large scale networks. As discussed in the next Section, there are many existing approaches to identify nodes with higher {\em centrality} indices, including {\em betweenness centrality} of particular interest to this paper. Existing exact deterministic algorithms to compute the nodes with high {\em betweenness centrality} typically employ Dijkstra's shortest path algorithm \cite{brandes:2001}, wherein they incorporate the dependency of a vertex on another single vertex, or the pair dependency to count the number of shortest paths for centrality evaluations. When executed in a graph $G(V, E)$, the running time is $O(|V||E|+|V|^2 log |V|)$.  In this paper, we aim to reduce the execution time by leveraging from the following key insight in modular social networks -  the impact of  {\em local betweenness centrality} indices in each local module of a large social network on the {\em global betweenness centrality} indices of the overall network. By a careful application of Dijkstra's shortest path algorithm among nodes in each local module (rather than on the global network) along with the sparse number of external edges (between modules) in such a manner that retains shortest path properties of the entire network, our algorithms to identify the {\em global betweenness central node} (i.e, the node in the overall network with the highest {\em betweenness centrality}), with significant savings in execution time. To the best of our knowledge, this is the first work that leverage modular properties of large scale social networks in efficiently computing {\em betweenness centrality}. 
\\The rest of the paper is organized as follows. In section II we discuss centrality measures as background. In section III we propose our approach and algorithm for computing betweenness centrality, while in section IV we discuss some characteristic of centrality relevant to our approach. Finally, in section V we discuss simulation results for our approach and conclude with future works in section VI.

\section{Background on Centrality Measures in Networks}

\label{sec:relat}
The {\em centrality} of a node in a social network can be quantified in multiple ways. Here we provide a brief description of four most representative  measures: {\em degree centrality}, {\em closeness centrality}, {\em betweenness centrality} and {\em eigenvector centrality}.

\begin{itemize}

\item \emph{Degree centrality} measures the total number of contacts incident upon a node, with importance in the design of forwarding algorithms. 
Evaluation of degree centrality is of the order of $O(|V|^2)$ over an underlying network graph $G(V, E)$, that reduces to the order of $O(|E|)$ for sparse graphs.
\item \emph{Eigenvector Centrality} measures the centrality of a vertex as the highest value from the normalized eigenvector that corresponds to the principle eigenvalue of the adjacency matrix for underlying application domain \cite{gould:1967}. Existing algorithms have a computational complexity of $O(|V|+|E|)$, although actual time depends on the spectral gap of the adjacency matrix.
\item \emph{Closeness centrality} \cite{ beauchamp:1965} measures how close a vertex is, to all other vertices in the graph. 
The closeness measure using the single source shortest path ($SSSP$) based exact algorithm of $Dijkstra$ \cite{coremen:2001} over $|V|$ source nodes is of the order of $O(|V||E|+|V|^2 log |V|)$.
\item \emph{Betweenness centrality} measures the importance of a node from the perspective of being located in the shortest paths that connect other peer nodes, thus captures both its load and significance in information flow over the network. Formally {\em betweenness centrality} of a node $v$ is the percentage of the number of shortest paths that node $v$ is part of between all pairs of nodes, over all possible shortest paths between them \cite{freeman1:1977}.  

Brandes \cite{brandes:2001} gives an exact algorithm  based on Dijkstra's shortest path algorithm \cite{coremen:2001} for computing betweenness centrality index of all nodes. Depending on the graph model, the exact algorithm takes time between $\theta(|V||E|)$ for unit edge weights and $\theta(|V||E|+|V|^2 log |V|)$ for general edge weights.
\end {itemize}

\section{Our Approach and Algorithms for Computing Betweenness Centrality}

\label{sec:III}

In this section, we present our algorithms for computing the {\em betweenness centrality} in large scale 
networks exhibiting modular topologies. In doing so, we first provide important definitions that will guide the rest of the paper. Then, we first present a modified version of Brandes algorithm for computing {\em betweenness centrality}. 
Subsequently, we discuss our approach for computing {\em betweenness centrality} taking advantage of modular topologies in social networks. Specifically, in our approach we introduce the local and global centrality concept that leverage the dense internal edges within module and sparse external edges among modules. 

\subsection{Definitions}

We consider a large scale social networks, denoted as $R(V, E)$, where $|V|$ denotes the number of vertices or actors or nodes, and $|E|$ denotes the number of edges. Based on the modular property of interest in this paper, we further consider the region $R$  divided into a set of independent modules $R_i(V_i, E_i)$, $R \cup_{i=1}^k R_i$, $\forall v_i \in V$, if  $v_i \in V_i \Rightarrow v_i \notin V\setminus V_i$. Let $w$ be the weight function over $E$, such that $w(e) > 0, e\in E$. The weight function can represent any parameter \emph{e.g.} connectivity, stress, cost, strength, demand. For an unweighted graph we define $w(e)=1$. Further, if there is no edge between a pair of vertices, then the edge weight is assumed to be $\infty$. As we mentioned earlier, we consider centrality indices based on shortest path weight measure over weighted graphs.
 \begin{defn}
\emph{Internal Edges}: In any module $R_i(V_i, E_i), \forall e_i(v_{k}, v_{l})$, such that $ v_{k}, v_{l}\in V_i$, then $e_i\in E_i$ and is called \emph{Internal edge} for module $R_i$.
\end{defn}

\begin{defn}
\emph{External Edges}: Across any pair of 
 modules $R_i(V_i, E_i), R_j(V_j, E_j), \forall e_{ij}(v_{k}, v_{l})$, such that $ v_{k}\in V_i, v_{l}\in V_j$, then $e_{ij}\in E$ is called \emph {External edge} for module $R_i$ and $R_j$.
\end{defn}

\begin{defn}
\emph{Local Centrality}: For any module  $R_i(V_i, E_i), V_i\subset V, E_i\subset E$, $LC(v_{ij})$ is the local betweenness centrality index associated with $v_{ij}$ computed over local topology for $R_i$, taking into account only internal edges.
The node $v_{ij}\in V_i$ having  maximum local centrality index $LC_i= max \{LC_{ij}\}$ is called local central node for that module.
\end{defn}

\begin{defn}
\emph{External Centrality}: For any set of modules over $R_i(V_i, E_i), V_i\subset V, E_i\subset E$, we define external betweenness centrality index ($EC$) for both nodes ($EC(v_{ij})$) and modules ($EC(R_i)$) as the number of times it acts as a connector or bridge across distinct modules.
\end{defn}

\begin{defn} \emph{Global Centrality}:
For region  $R=(V, E)$, global centrality of a vertex $v_{ij}$ takes into consideration all (both internal and external ) edges over region $R$. If $GC_{ij}$ is the global centrality index of vertex $v_{ij}$, then global centrality $GC=max\{GC_{ij}\}$ and the corresponding node $v_{ij}$ is called global central node.
\end{defn}

\begin{defn}\emph{Global Central Module}:
For any module $R_i$, we define external centrality index for $R_i$ as $EC(R_i)= \sum_{v_{mn}} EC(v_{mn})$, where $v_{mn}$ is the set of nodes in $V_i$ with incident external edges. The module $R$ with maximum external centrality index $EC(R_i)= MAX (EC(R_i)), \forall R_i \subset R$ is called the global central module.
\end{defn}

\subsection{Problem Definition}
Given a set of modules $R_i(V_i, E_i)$ with respective module boundaries we evaluate local and global betweenness centrality indices and compute local and global central node, and global central module.

\subsection{Adaptation of Brandes Algorithm}
The following exact algorithm is used to compute centrality indices of a set of nodes and then determine the node with highest centrality index value. Let $\{v_{ext}\}$ represent the set of nodes with external edges incident to them. We keep record of the shortest path weights and the number of vertices on the path from every node in $V_i$ to every node in corresponding $\{v_{ext}\}$, which is used in computing global centrality indices.


\begin{algorithm}
\caption{Calculate $BC(R)$}
\begin{algorithmic}
\REQUIRE $|V|=n$ nodes in network $R$ and the edges connecting them
\ENSURE Betweenness centrality indices in $R$

\FOR{$i \leftarrow 1$ to $|V|$}
\STATE $l_i\leftarrow 0$, $V^T_i\leftarrow v_i$, $E^T_i\leftarrow \phi$
\STATE Find smallest edge $l_e=(v_i, v_j)$
\STATE $l_{ij}\leftarrow l_i+l_e$
\STATE  $E^T_i\leftarrow E^T_i\cup (v_i, v_j)$,
$V^T_i\leftarrow  V^T_i\cup \{ v_j\}$,
$LC(T_{v}) \leftarrow 0$, $v\in V^T_i$
\STATE $S_{st}(v_i)=0, S_{st}(v_j)=0$
\STATE $D_s(v_i)=S_{st}(v_i)/S_{st}=0$ \\and $D_s(v_j)=S_{st}(v_j)/S_{st}=0$
\STATE $P(v_j)=1$, $v_j\neq v_{ext}$
\ENDFOR

\WHILE{$|V^T_i| \neq n$}
\FOR{$i\leftarrow 1$ to $n$}
\STATE find $v_k$ such that $MIN((l_{i\leadsto j}+l_{jk}), l_{ik}) $, $l_{i\leadsto j}=w(v_i\leadsto v_j)$, $l_{jk}=w(v_j, v_k)$, $l_{ik}=w(v_i, v_k)$, $\forall v_j \in V^T_i$, $(v_i,v_k), (v_j,v_k)\notin E^T_i, v_k \notin V^T_i$.
\STATE Count $S_{st}(v_j)$, $\forall v_j $ in $v_i\leadsto v_k$ and $v_j \neq v_i \neq v_k$
\STATE Count $P(v_k)$, where $v_k\neq v_{ext}$
\STATE Compute $ D_s(v_i)=D_s(v_i)+S_{st}(v_i)/S_{st}$, $ v_i, t\in V^T_i$ for each node $s$ corresponding to each list.

\ENDFOR


\ENDWHILE

\STATE Compute $BC(T_{v}) = \sum_{ s, t} D_s(v), \forall s,  s\in V ,\forall t, t\in V\setminus s$, and $v\in V$
\STATE $BC_{max}=MAX(BC(T_{v}))$
\end{algorithmic}
\end{algorithm}

In this algorithm each node in the region is initialized as the start vertex $s$ and shortest path from that node to all other vertices in the region is computed. For any vertex $v_i$, distance to itself is $0$. Initially, the network $R$ with $|V|$ nodes is initialized as $|V|-$lists (pockets) of one node each acting as starting node. Let $V^T_i$ represent the set of travelled vertices. The weighted path $(v_i\leadsto v_k)$ or weighted edge $(v_i, v_k)$ is denoted as $l_{i\leadsto k}$ or $l_{ik}$, respectively. The smallest weighted edge $l_{ij}=w(v_i,v_j)$ is selected from each of these vertices ($\{v_i\}$); the edge is added to the list of travelled edges $E^T_i$ and the node at the other end  of the edge is added to
$V^T_i$. In each iteration a new node is added to every list corresponding to each of the $|V|$-vertices. The selection of new node $v_k\notin V^T_i$ is based on minimum weighted path $w(v_i\leadsto v_k)= l_{i\leadsto k}$, where $v_i$ is the source node in the respective list. This minimum $l_{i\leadsto k}$ is computed over the path $(l_{i\leadsto j} + l_{jk})$, and the edge $l_{ik}$. When there is only one vertex in the list $V^T_i$ ($i=j$), $l_{ii}=0$, the new node $v_k$ with minimum direct edge weight $w(v_i, v_k)=l_{ik}$ is added to the list. In case of $1<|V^T_i|<|V|$, $l_{i\leadsto k}$ is computed over all $v_j \in V^T_i$ and $l_{ik}$. Each time a node $v_k$ is added to the list, the number of shortest paths going through any vertex $v$, $S_{st}(v)$ for all nodes in the path $v_i\leadsto v_k$ is increased by one.

In contrast to the pair dependency sum in Dijkstra's algorithm for finding \emph{betweenness centrality}, the Brandes approach considers the dependency of a node on another node for which betweenness centrality is being evaluated. Let $v$ be the vertex for which betweenness centrality is being evaluated, and $s$ be the starting node then the presence of $v$ in all $s\leadsto t$ path contributes to the dependency of $s$ on $v$. This is represented as $D_s(v)$. Further, to compute betweenness centrality index for $v$, the dependency on $v$ is computed $\forall s,t  \in V$, where each $s$ corresponds to a list and $t\in V\setminus s$.

When shortest path to each of the $|V|$ vertices is computed, 
the algorithm stops, computing the betweenness centrality index for each vertex $v$, $BC(T_{v})$ and then evaluating the node with highest centrality index $BC_{max}$.

The \emph{Correctness of the algorithm depends on the } the fact that in each iteration nodes with least weighted path are added to $V^T_i$. And that $BC(v)$ is evaluated over all shortest paths through $v$.

The complexity of \emph{Algorithm I} is of the order of $O(|V||E|+|V|^2 log |V|)$ \cite{coremen:2001, brandes:2001}.

In the following subsection we present in detail how the above algorithm is used in our proposed framework.

\subsection{Our Algorithm}
In our approach, we define global centrality indices for any vertex $v_{ij}$ as $GC(v_{ij})=LC(v_{ij})+EC(v_{ij})$, where $EC(v_{ij})$ is the additional centrality index due to external edges, where $EC(v_{ij})=S_{st}(v_{ij})/S_{st}, s\neq t\neq v_{ij}$, $S_{st}$ is the number of shortest paths between nodes $s$ and $t$, and $S_{st}(v_{ij})$ is the number of shortest paths between nodes $s$ and $t$ going through $v_{ij}$, $s\in V_i \Rightarrow t\in V\setminus V_i$. And  the node $v_{ij}$ with maximum global centrality index $GC=max\{GC(v_{ij})\}$ is called global central node over the network.

In our proposed \emph{Algorithm II} we consider internal edge weights while computing external edge based centrality indices $EC(v_{ij})$. It is because in application domains such as smart grid and disease transmission, internal edges have role beyond their own module; here, depending on which node in the module is connected to an external edge, the corresponding internal edges have effect on path weight to other nodes within the module. For example, between any pair of interacting individuals in a disease propagation network, an individual's degree of immunity to a disease comes into picture. Further, in our example academic network, if nodes inside a module are not equally comfortable in reaching a node in $\{v_{ext}\}$ for collaboration (or communication) with nodes in other modules, then internal edges weighted as degree of comfortability should be taken into consideration.  

In contrast, in our \emph{Algorithm III} we do not take into consideration the internal edge weights while computing external edge based centrality $EC(v_{ij})$. It is because, in some applications \emph{e.g.} physical Internet domain, and some cases of social network domain internal edges does not contribute much when we consider external edges among regions and compute global centrality. For example: when a packet comes through external edge from one region to another, within the LAN it does not matter to which specific node it is destined to as, that does not incur much cost; similarly in case of the earlier example of academic network  we stated above, if one or more person from one lab has contact with other labs then all the persons in the corresponding labs are  reachable to each other via those nodes, without much internal cost.

Below we consider how external centrality indices is computed for both nodes and modules using internal and external edge weights, and how global centrality indices are evaluated. We hereby present  our \emph{Algorithm II}.

We use \emph{Algorithm I} over modules $R_i$ to obtain local centrality indices of nodes, keep record of their shortest path to vertices in $v_{ext}$ and the number of nodes on the corresponding shortest path. Over each  module $R_i$ we know the set of vertices $v_{ext}$ with incident external edges $e_{ext}$, weight of the external edges $w(e_{ext})$, and the shortest path weight from vertices within that module to the vertices in  $\{v_{ext}\}$. In an attempt to find the shortest path to nodes across modules, we first find the egress points to outer modules. For any module $R_i$ the minimum weighted egress edge is computed as the $MIN(l_{i\leadsto k^i_{ext}}+l_{k^i_{ext}k^j_{ext}})$, where $l_{i\leadsto k^i_{ext}}=w(v_i\leadsto v_{k_{ext}})$ is the path weight from node $v_i$ to $v_k\in \{v_{ext}\},  \{v_{ext}\} \subset V_i$ in the same module, $l_{k^i_{ext} k^j_{ext}}=w(v_{k^i_{ext}}, v_{k^j_{ext}})$ with $v_{k^i_{ext}}$ a node in $R_i$ with incident external edge and  $v_{k^j_{ext}}$ is the other end of the external edge in module $R_j$, $MIN$ is the minimum weight function. In $R_i$ this evaluation over all external edges and vertices $V_i$ gives the egress path for all the vertices from the module.  Since the set of external edges are sparse, a subset of vertices from any module follow the same egress edge. The above computation is repeated over all the modules, creating a partition of vertices over each module. The subset of vertices following one egress path are recorded, so is the set of travelled vertices. Once the egress path for vertices from each module is known, the set of vertices in other modules, or external edges to other modules that they follow is determined by $MIN((l_{i\leadsto k^j_{ext}}+ l_{k^j_{ext}\leadsto v_j}), (l_{i\leadsto k^j_{ext}}+ l_{ k^j_{ext} \leadsto k^m_{ext}}), (l_{i\leadsto k^i_{ext}} + l_{k^i_{ext}\leadsto k^j_{ext}} + l_{k^j_{ext}\leadsto v_j}))$ where, $ l_{k^j_{ext}\leadsto v_j}$ is the shortest path from $v_{k_{ext}}\in V_j$ to any vertex $v_j\in V_j$, $l_{ k^j_{ext} \leadsto k^m_{ext}}$ is the shortest path between two vertices with incident external edges in distinct modules, $(l_{i\leadsto k^i_{ext}} + l_{k^i_{ext}\leadsto k^j_{ext}} + l_{k^j_{ext}\leadsto v_j})$ is the shortest path from $v_i$ to $v_{k_{ext}}\in V_i$ and following the shortest egress path to another module $R_j$ and the set of vertices in module $R_j$. The set of vertices covered in any path are kept in the corresponding list for vertex $v_i\in V$ called $V^T_i$. When $V^T_i==V\setminus V_i$, the algorithm stops repeating for the vertex $v_i$, where $v_i\in V_i$. For every traversed shortest path to other module, the betweenness centrality of a vertex is computed as the measure of the dependency of rest of the vertices on it. This gives the  external centrality measure, $EC(v_i)$.  The computation of the metric $EC(R_i)=\sum EC(v_i), v_i\in \{v_{ext}\}, \{v_{ext}\}\subset V_i$ gives the influence that the external vertices of a module $R_i$ have, across all modules. Further, evaluating  $MAX(EC(R_i))$ helps us finding the module with highest influence index, hence is called the global central module. At the beginning, once \emph{Algorithm I} is run over each module, each node has an associated local centrality index. We define intermediate centrality index for each vertex as $IC(v_i)$, initialized as $IC(v_i)=LC(v_i)$. 
During the evaluation of $EC(v_i)$, the intermediate centrality becomes $IC(v_i)+EC(v_i)$. 
By taking into consideration the dependency of all vertices on $v_i$ during their corresponding shortest paths, the centrality index $IC(v_i)+EC(v_i)$ represents the global centrality index for vertex $v_i$. The node with maximum value of $IC(v_i)+EC(v_i)$ becomes the global central node. The initial subset of vertices following an egress path, are most likely to follow the same path, adding external vertices and internal vertices of other modules to the respective lists for every element in the subset. It is because any node follows an egress path iff its path weight to the egress point and weight of the egress path is least among all possible alternate paths.

\begin{algorithm}
\caption{Calculate $LC(R_i)$ and $GC(R)$}
\begin{algorithmic}
\REQUIRE $R_i$s with corresponding nodes $|V_i|$, number of modules: k and edge set $E$ over network $R$
\ENSURE Local betweenness centrality over module $R_i$:  $LC(R_i)$ and Global betweenness centrality  in $R$: ($GC(R)$)
\FOR{$r\leftarrow 1$ to $k$}
\STATE use \emph{Algorithm I} to compute $LC(R_r)$, using only edges $(v_i, v_j)\in R_r$
\STATE  Initialize $IC(v_i)=LC(v_i)$
\ENDFOR

\FOR{$r\leftarrow 1$ to $k$}
\WHILE{$V^T_i \neq V\setminus V_i$}

\STATE  Find egress edge for all $v_i\in V_i$. Compute $MIN(l_{i\leadsto k^i_{ext}}+l_{k^i_{ext}k^j_{ext}})$.  
\STATE Find the minimum cost path across modules. Compute $MIN((l_{i\leadsto k^j_{ext}}+ l_{k^j_{ext}\leadsto v_j}), (l_{i\leadsto k^j_{ext}}+ l_{ k^j_{ext} \leadsto k^m_{ext}}), (l_{i\leadsto k^i_{ext}} + l_{k^i_{ext}\leadsto k^j_{ext}} + l_{k^j_{ext}\leadsto v_j}))$


\STATE Count $S_{st} (v_j)$, $\forall v_j $ in $v_i\leadsto v_k$ and $v_j \neq v_i \neq v_k$
\STATE Compute $ D_s(v_i)=D_s(v_i)+S_{st}(v_i)/S_{st}$, $ v_i, t\in V^T_i$ for each node $s$ corresponding to each list.


\ENDWHILE

\ENDFOR
\STATE Compute $BC(T_{v}) = \sum_{ s, t} D_s(v), \forall s,  s\in V ,\forall t, t\in V\setminus s$, and $v\in V$
\STATE Compute $EC(R_i)=\sum EC(v_i), v_i\in \{v_{ext}\}, \{v_{ext}\}\subset V_i$
\STATE Compute global central module $R_i$, $GC(R_i)=MAX(EC(R_i))$
\STATE Compute global central node $v_i$, $GC(v_i)= MAX(BC(T_{v}) ) $
\RETURN$GC ( R )$, $GC(v_i)$

\end{algorithmic}
\end{algorithm}

Next, we consider the procedure for evaluating centrality in an application domain, where internal edges within a module does not contribute to external centrality evaluation among modules. We present it in \emph{Algorithm III}. As defined earlier, the set of external vertices in a module represented as $v_{ext}$ with incident external edges $e_{ext}$. Once the local centrality indices are evaluated over each module using algorithm I, the external centrality indices are evaluated using only external edges over modules. Each module can have multiple external edges to other modules. Let's visualize the modules as virtual nodes with set of incident external edges, over which shortest path among modules is evaluated. For each module we consider a list of travelled modules $R^T_i$ that is initialized to the respective module itself, $R_i$. From each module the external edge with $MIN(R_{\{i_{ext}\}})$ is followed, where $R_{\{i_{ext}\}}$ is the set of external edges from module $R_i$, $MIN$ is the minimum weight function; the corresponding module at the other end of the external edge is added to the respective list $R^T_i$. Let the number of modules be $\#(R_i)$. When $2\leq|R^T_i|\leq \#(R_i)$, each module $R_i$ evaluates the metric: $MIN((R_{i_{ext}k_{ext}}), (R_{i_{ext}\leadsto j_{ext}}+R_{j_{ext}k_{ext}}))$, adding the corresponding module to their respective list $R^T_i$, where $R_{i_{ext}k_{ext}}$ is the external edge between modules $R_i$ and $R_k$ adding module $R_k$ to the list, $R_{i_{ext}\leadsto j_{ext}}$ is the shortest path between modules $R_i$ and $R_j$ along a set of external edges with $R_j \in R^T_i$, $R_{j_{ext}k_{ext}}$ is the external edge between modules $R_j$ and $R_k$ with $R_j\in R^T_i, R_k\notin R^T_i$. For each $R_i$ its shortest path $R_{i_{ext}\leadsto k_{ext}}$ through all $R_j$ increases dependency of $R_i$ on $R_j$, that is increases the betweenness centrality index for all such $R_j$. When every list corresponding to respective $R_i$, $R^T_i==\#(R_i)$, the corresponding centrality index $EC(R_i)$ gives the influence of a module across all modules. The module with maximum index value for $EC(R_i)$ gives the globally central module.
The external centrality index for any vertex $v_i$ is evaluated as $EC(v_i)=O(k\times l)$, where $k$ is the number of nodes in one module and $l$ is the number of nodes in modules other than that module and they communicate via $v_i$. When $v_i\in \{v_{i_{ext}}\}$, the external vertices with incident external edges that are actually followed as a connector among modules are only considered for computing $EC(v_i)$. The global centrality index for any vertex  $v_i$ is given by $IC(v_i)+EC(v_i)$, $IC(v_i)$ is the existing intermediate centrality index of the vertex $v_i$ initialized as $IC(v_i)=LC(v_i)$.
Further, an weighted version of evaluation for $EC(v_i), v_i\in \{v_{_{ext}}\}$ can be done as  $EC(v_i)=(|k||l| w(e_{i_{ext}})/(\sum w(e_{i_{ext}})))$, where $w(e_{i_{ext}})$ is the weight of a single external edge incident on $v_i$ and $(\sum w(e_{i_{ext}}))$ is the sum of all external edge weights incident on $v_i$; $k, l$ as defined earlier. The node with maximum value of $IC(v_i)+EC(v_i)$ becomes the global central node.


\begin{algorithm}
\caption{Calculate $LC(R_i)$ and $GC(R)$}
\begin{algorithmic}
\REQUIRE $R_i$s with corresponding nodes $|V_i|$, number of regions: k and edge set $E$ over network $R$
\ENSURE Local betweenness centrality over module $R_i$:  $LC(R_i)$ and Global betweenness centrality  in $R$: ($GC(R)$)

\FOR{$r\leftarrow 1$ to $k$}

\STATE use \emph{Algorithm I} to compute $LC(R_r)$, using only edges $(v_i, v_j)\in R_r$
\STATE  Initialize $IC(v_i)=LC(v_i)$
\STATE  From each module consider egress edge: $MIN(R_{\{i_{ext}\}})$
\STATE Initialize $R^T_i$.

\ENDFOR

\WHILE{$R^T_i\neq \#(R_i)$}
\FOR{$r\leftarrow 1$ to $k$}

\STATE Find minimum cost path across modules: Compute $MIN((R_{i_{ext}k_{ext}}), (R_{i_{ext}\leadsto j_{ext}}+R_{j_{ext}k_{ext}}))$

\STATE Count $S_{st} (v_j)$, $\forall v_j $ in $v_i\leadsto v_k$ and $v_j \neq v_i \neq v_k$
\STATE Compute $ D_s(v_i)=D_s(v_i)+S_{st}(v_i)/S_{st}$, $ v_i, t\in V^T_i$ for each node $s$ corresponding to each list.

\ENDFOR
\ENDWHILE

\STATE Compute $BC(T_{v}) = \sum_{ s, t} D_s(v), \forall s,  s\in V ,\forall t, t\in V\setminus s$, and $v\in V$
\STATE Compute $EC(R_i)=\sum EC(v_i), v_i\in \{v_{ext}\}, \{v_{ext}\}\subset V_i$
\STATE Compute global central module $R_i$, $GC(R_i)=MAX(EC(R_i))$
\STATE Compute global central node $v_i$, $GC(v_i)= MAX(BC(T_{v}) ) $
\RETURN$GC ( R )$, $GC(v_i)$

\end{algorithmic}
\end{algorithm}

In both these algorithms the modules under network $R$ is represented as $R_i$. In algorithm III the internal path cost to the vertex with incident external edge is not taken into consideration while computing centrality over external edges, where as in the second algorithm this is an intrinsic part of the computation. Further, whether we take internal edge weights into consideration, also depends on the diameter of the module.

\begin{proposition}
\emph{The central node with highest betweenness centrality index is preserved as the node with maximum global centrality index in our first approach, \emph{Algorithm II}.}\\
{\bf Proof:}\\
In our first approach that is Algorithm II we consider global betweenness centrality index of any vertex as $GC(v_i)=LC(v_i)+EC(v_i)$. The network domain can be visualized to be partitioned as $V_i$ and $V\setminus V_i, v_i\in V_i$. Since, centrality is additive, the betweenness centrality of any vertex can be considered as the dependency of the nodes in $V_i$ on $v_i$ and the nodes in $V\setminus V_i$ on $v_i$. In the first case we record the dependency of nodes in $V_i$ on $v_i$ as the local centrality index $LC(v_i)$ and is evaluated according to shortest path measure. In the second case we consider the effect of $V\setminus V_i$ on $v_i$, as the dependency of rest of the modules on $v_i$, in obtaining shortest paths across modules. In computing this not only do we consider shortest path across modules through external edges, but also take into account the shortest path of nodes within modules to their respective external vertices for egress path. Here, the centrality evaluation involving external edges gives us the dependency on a node, while traversing shortest path from nodes within one module to nodes across other modules, giving us $EC(v_i)$. Thus, $GC(v_i)$ takes into account dependency of all vertices on $v_i$ while evaluating betweenness centrality index of $v_i$, and preserves the betweenness centrality index using existing deterministic approach. Thus, the global central node with highest index value is also preserved.\\
Further, our algorithm II correctly evaluates the shortest path based betweenness centrality index as, at each phase its choice of nodes to $V^T_i$ depends on the minimum weighted path from the starting vertex to an external vertex in the module, to external vertex across modules following external edges and from external vertex to other nodes in that module by evaluating the metric $MIN(l_{i\leadsto k^i_{ext}}+l_{k^i_{ext}k^j_{ext}})$ and $MIN((l_{i\leadsto k^j_{ext}}+ l_{k^j_{ext}\leadsto v_j}), (l_{i\leadsto k^j_{ext}}+ l_{ k^j_{ext} \leadsto k^m_{ext}}), (l_{i\leadsto k^i_{ext}} + l_{k^i_{ext}\leadsto k^j_{ext}} + l_{k^j_{ext}\leadsto v_j}))$. So, any external edge (hence external vertex) or internal nodes not on the shortest route are ever aded to the $V^T_i$. Thus, preserving the shortest path measure in evaluating betweenness centrality index.

 \end{proposition}

 \begin{proposition}
\emph{The correctness of \emph{Algorithm III} depends on the fact that the global central node is preserved under the constraint that densely connected vertices does not account for shortest path in computation of $EC(v_i)$.} \\
Here, our algorithm evaluates global betweenness centrality index as $LC(v_i)+EC(v_i)$, where $LC(v_i)$ takes into consideration the effect of dense internal edges on a node and $EC(v_i)$ considers the effect of external edges only, as explained earlier. So, $EC(v_i)$ considers the shortest paths across modules, adding the effect of respective external vertices and internal vertices, though the path from a node within a module to the corresponding external vertex and vice versa are not based on shortest paths. Here, the centrality index of nodes from existing computation is not preserved, as nodes without being on the shorter path can have higher dependency measure. Thus, the global central node is not preserved in algorithm III, though in a constrained evaluation of the existing algorithm nullifying the effect of dense internal edge weights gives us centrality indices evaluated here.

 \end{proposition}

In \emph{Algorithm II} the centrality evaluation is based on the number of external edges from a module, number of modules, number of nodes in each module and the number of internal edges. Let $m$ be the number of nodes in each module, $n$ be the number of modules, $e'$ be the number of external edges and $e$ be the number of internal edges. Running algorithm I on each module to evaluate local centrality will take time $O(me+m^2 log m)$. Time it takes over all modules is given by $O(n(me+m^2 log m))$. In order to compute external centrality indices the algorithm depends on the number of external edges, number of nodes in modules, shortest path from nodes to $v_{ext}$ in any module (this value is precomputed during local centrality evaluation). Further in any module $e' << m$, due to the sparsity of external edges that further gives to a partition of vertices in each module over egress path. In evaluating external centrality indices the computation time becomes $O(n(me'+m^2 log m))$. Thus, computation time to evaluate global centrality indices given by $O(n(me+m^2 log m))+ O(n(me'+m^2 log m))=O(n(me+m^2 log m))$. Here, having large number of nodes in any module does not give any positive effect on computational cost. Similarly, is the case for tiny number of nodes in each module with large number of modules. We thus consider an ideal case say $n=\sqrt |V|$. Given the total number of nodes in the application domain as $|V|$, the average number of nodes per module is $\sqrt |V|$. With dense internal edge set, lets consider $e=\Mycomb [\sqrt |V|] {2}=O(|V|)$ and sparse external edge set $e'=\frac{\Mycomb [\sqrt |V|] {2}} 2=O(|V|)$. Thus incurring total computation cost of $O(|V|^2+|V|^{1.5} log |V|)$.

Further, \emph{Algorithm III} takes into account internal path cost while computing the local centrality, but considers only shortest path over external edges while computing $EC(v_i)$. Here, during computation of local centrality the computation time is of the order of $O(n(me+m^2 log m))$. During the evaluation of external centrality the computation time depends only on the number of external edges and the number of modules thus, giving evaluation cost of the order of $O(ne'+n^2 log n)$.  This gives us incurred computational cost of the order of $O(n(me+m^2 log m)+ ne'+n^2 log n)=O(n(me+m^2 log m))$, this for the ideal case gives us $O(|V|^2+|V|^{1.5} log |V|)$.

Incase of arbitrary module size and number of modules, the evaluation of local centrality and the recording of the shortest paths to external vertices can be done during module construction phase and can be considered as preprocessing cost. This helps reduce the computational cost to $O(|V|^{1.5}+|V| log |V|)$.

In the following section we discuss interesting centrality characteristics relevant to our work.
\section{Discussion}
In our work we discussed ways to evaluate local central node and global central node, so it is interesting to examine if a local central node ultimately become global central node. A local central node say $v_i$ has centrality index $LC(v_i)$. To be a candidate for global central node from its module $R_i$ it needs to have maximum global centrality index ($LC(v_i)+EC(v_i)$). The external edges to $R_i$ can be incident to local central node or, a combination of local central node and local non-central node or, to local non-central node.
When incident to only local central node, its centrality index becomes $LC(v_i)+O(|V_i|\times \sum_{j\in R\setminus R_i} |V_j|  )$, where $|V_j|$ is the number of vertices in other modules that have external shortest path to $R_i$ through $v_i$. Here, local central node becomes candidate for global central node. In order to weigh the possibility in last two cases, lets consider the worst case where, non of the external edges are incident to the local central node. Here, it needs to investigate: whether the shortest path to the local non-central node with incident external edges exist via local central node or not. When there is shortest path through local central nodes, its centrality index increases by at most $EC(v_i)=O(k\times \sum_{j\in R\setminus R_i} |V_j|  )$, where $k$ is the set of nodes in $R_i$ dependent on $v_i$. Where as, the local-non central node has the highest $EC$ index. Here, depending on who has higher global centrality index $LC(u)+EC(u)$, $u$ vertex for which centrality is evaluated, either the local central node or the local non-central node becomes the candidate for global central node from module $R_i$. Similar is the argument for the last possibility. This justifies that local central node may not be the global central node.

We further consider the possibility of global central node being part of global central module.
For a node $v_i$ to be global central node its centrality index $LC(v_i)+EC(v_i)$ is maximum over $V$.
Consider the global central module. Lets consider the worst case with $LC(v_i)=0$ but, centrality index due to external edges in module $R_i$, $EC(R_i) > EC(R_j)$, $\forall  R_j\in R\setminus R_i$ thus making it the global central module. In other words,  $\sum EC(v_i)=EC(R_i)$, where vertices ($v_i$) have incident external edges. Consider module $R_k$ with $EC(R_k) < EC(R_i)$ and $\exists v_k \in R_k$ where, $LC_k(v_k)=MAX (LC_k(v_i)), v_i, v_k \in R_k$ and $v_k\neq v_i$. Let $LC_k(v_k) - LC_i(v_i) >EC(v_i) - EC(v_k)$, where $LC_i(v_i)$ is local centrality index of $v_i$ in community $R_i$ and $EC(v_i)$ is the corresponding external edge based centrality for vertex $v_i$ in module $R_i$. Thus, even if module $R_i$ is the global central module, the global central node is $v_k, v_k\notin R_i$.

\section{Simulation}
\label{sim}

In order to evaluate our approach we consider a synthetic graph with $5000$ nodes. For any given network with $|V|$ nodes and undirected edges among them, the maximum possible number of edge are given by $C^{|V|}_2=(|V|^2-|V|)/2$, where as, the minimum  and average number of edges are known to be of $|V|-1$ and $(|V|^2-|V|)/4$. In a sparse graph though number of edges can never exceed $(|V|^2-|V|)/4$ and can be at least of $|V|-1$ numbers to maintain connectivity. In our weighted synthetic graph we assign edge weights in the range of $10$ to $50$. In order to examine the effectiveness of the proposed modular structure  we compare it with the existing Brandes approach where, the complete network region with $|V|$ nodes is considered instead of a set of modules. For both cases we consider average number of edges  and the number of modules for our approach as the $\sqrt |V|$, where $|V|$ is the number of nodes considered in Brandes  approach and each module has $\sqrt |V|$ nodes. We plot the variation of computation time (in seconds) for both cases in Figure 2, as a bar chart, where our approach fairs better and more so with higher the number of nodes.
Lastly we investigate how for a given fixed size of nodes in the network graph, two different module formation effects the computational cost. We vary our number of nodes from $1000$ to $5000$. For module formation in one case we consider $\sqrt |V|$ number of modules, with $\sqrt |V|$ nodes per module; in the other case we consider approximately one hundredth of the number of nodes as the number of modules. We plot the variation in computational cost for both the cases in Figure 2.

All our simulations are done in 2.3 GHz Intel core i5 processor with 4 GB RAM.

\begin{figure*}[!htb]
    \centering
  \hspace{4mm}  \includegraphics[scale=.7]{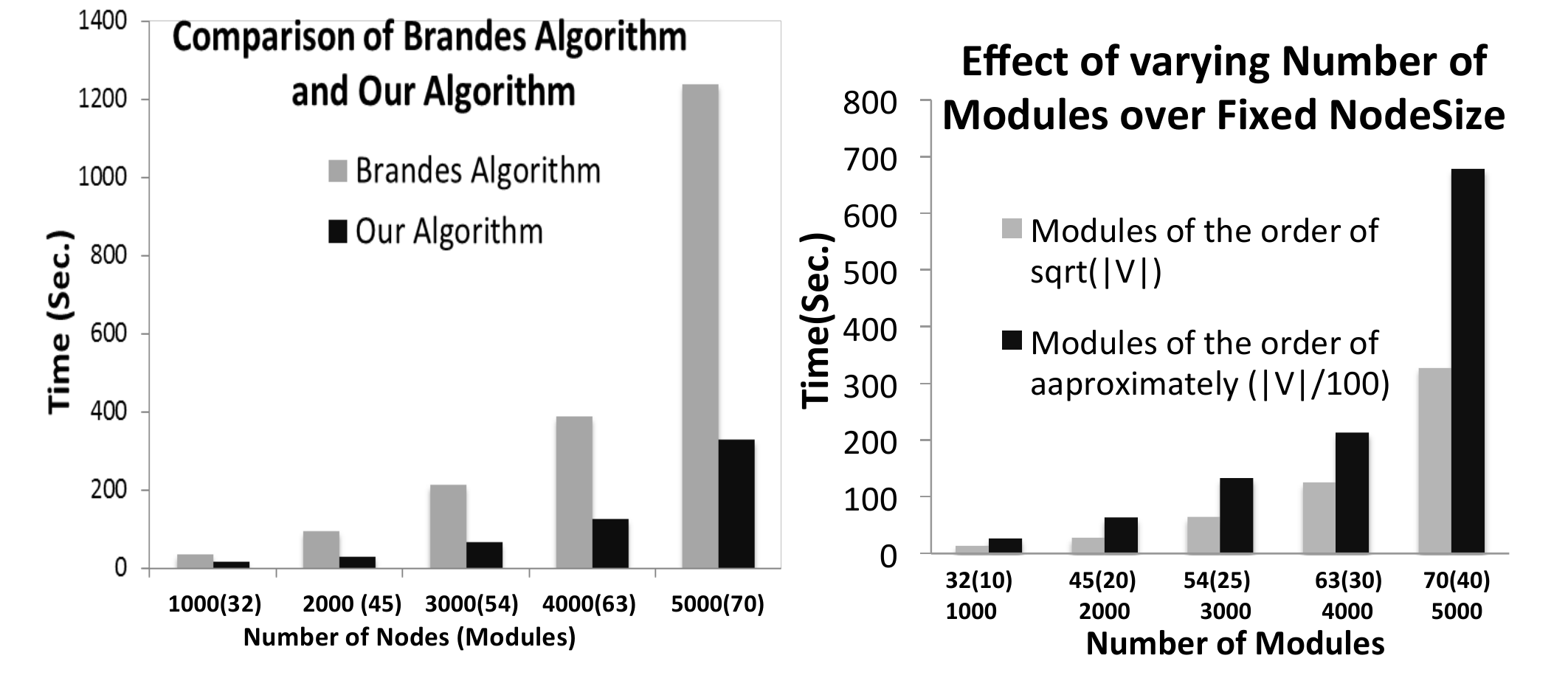}
   \caption{Simulation results of our Algorithm}
    \label{fig:hist}
\end{figure*}

\section{Conclusion}

In this paper we propose a modular framework for evaluating centrality by considering the density and sparsity of internal and external edges respectively. 
For an application network with $|V|$ nodes, in the ideal case the computational cost for betweenness centrality indices is  bounded by $O(|V|^2+|V|^{1.5} log |V|)$.
We also prove that a local central node belonging to the global central module may not becomes the global central node. Further, in this approach we also get the global central module that is most influential across modules.

In addition, the modular structure helps in finding a module that has the most influential nodes across modules. In many applications these nodes and hence the global central module plays the important role as information spreaders and are of far more significance than the global central node itself. To us, consideration of betweenness centrality over the whole application domain to find the most influential or critical nodes, is biased by the dense internal connections in the module in which they belong to. Since, these modules have dense internal edges, anyway information is going to spread, but what is most critical is the central nodes across the modules, where connections are sparse and they actually help diffuse information globally. Thus, in contrast to the existing single notion of global central node, our framework helps distinguish between global central nodes with maximum global central index and nodes with highest external edge based centrality. These, second type of nodes acts as the major connectors across modules.
In our future work we are going to characterize the effect of and evaluate these new externally central nodes.

In this work we considered betweenness as our centrality measure, but our approach is equally applicable to closeness centrality measure too, where we only need to consider summation of shortest paths from each vertex to every other vertex. The intrinsic modular structure implies that within a module most nodes are going to have better (higher) closeness centrality index attributed to the dense internal edges. Further, with modular structure we are going to have savings in computational cost as in case of betweenness centrality. In contrast to the existing approaches, with our approach we can distinguish nodes that are closest to nodes across the modules, that act as significant nodes(actors) in reaching out across modules faster. And these nodes are not always same as the nodes with highest closeness centrality measure, using the existing approach.

Here we propose the framework for static structure without any consideration of temporal effect. We would also like to consider the temporal variation, its effect in the proposed framework and centrality evaluation; and consider a multi-tier approach where each tier of the hierarchy will encompass  a set of modules as virtual nodes and hence incorporate the dynamism  \cite{wen:2013}.



\bibliographystyle{IEEEtran}
\bibliography{MyThesis}

\begin{thebibliography}{10}
\providecommand{\url}[1]{#1}
\csname url@samestyle\endcsname
\providecommand{\newblock}{\relax}
\providecommand{\bibinfo}[2]{#2}
\providecommand{\BIBentrySTDinterwordspacing}{\spaceskip=0pt\relax}
\providecommand{\BIBentryALTinterwordstretchfactor}{4}
\providecommand{\BIBentryALTinterwordspacing}{\spaceskip=\fontdimen2\font plus
\BIBentryALTinterwordstretchfactor\fontdimen3\font minus
  \fontdimen4\font\relax}
\providecommand{\BIBforeignlanguage}[2]{{%
\expandafter\ifx\csname l@#1\endcsname\relax
\typeout{** WARNING: IEEEtran.bst: No hyphenation pattern has been}%
\typeout{** loaded for the language `#1'. Using the pattern for}%
\typeout{** the default language instead.}%
\else
\language=\csname l@#1\endcsname
\fi
#2}}
\providecommand{\BIBdecl}{\relax}
\BIBdecl

\bibitem{michalski:2011}
R.~Michalski and S.~Palus, \emph{Modelling Social Network Evolution}.\hskip 1em
  plus 0.5em minus 0.4em\relax Springer Verlag, 2011, vol. 6984.

\bibitem{viswanath:2011}
B.~Viswanath, A.~Post, and K.~P. Gummadi, ``An analysis of social network-based
  sybil defenses,'' in \emph{Proceedings of the ACM SIGCOMM}, ACM, Ed.,
  vol.~40, 2010, pp. 363--374.

\bibitem{olsson:2004}
P.~Olsson and C.~Folke, ``Adaptive comanagement for building resilience in
  social-ecological systems,'' \emph{Environmental Management}, vol.~34, no.~1,
  pp. 75--90, July 2004.

\bibitem{gross:2005}
R.~Gross and A.~A., ``Information revealation and privacy in online social
  networks,'' in \emph{Proceedings of the 2005 ACM Workshop on Privacy in the
  Electronic Society}, ser. WPES '05'.\hskip 1em plus 0.5em minus 0.4em\relax
  ACM, 2005, pp. 71--80.

\bibitem{beach:2009}
A.~Beach, M.~Gartrell, and H.~Richard, ``Solutions to security and privacy
  issues in mobile social networking,'' in \emph{Computatinal Science and
  Engineering, 2009. CSE' 09. International Conference on}, vol.~4.\hskip 1em
  plus 0.5em minus 0.4em\relax IEEE, 2009, pp. 1036--1042.

\bibitem{bavelas:1948}
A.~Bavelas, ``A mathematical model for group structures,'' in \emph{Human
  Organization}, vol.~7, 1948, pp. 16--30.

\bibitem{leavitt:1951}
H.~J. Leavitt, ``Some effects of communication patterns on group performance,''
  \emph{Journal of Abnormal and Social Psychology}, vol.~46, pp. 38--50, 1951.

\bibitem{watts1999networks}
D.~J. Watts, ``Networks, dynamics, and the small world phenomenon,''
  \emph{American Journal of Sociology}, vol. 105, no.~2, pp. 493--527, 1999.

\bibitem{hanneman2005introduction}
R.~A. Hanneman and M.~Riddle, ``Introduction to social network methods,''
  University of California Riverside, 2005.

\bibitem{mishra2007clustering}
N.~Mishra, R.~Schreiber, I.~Stanton, and R.~E. Tarjan, ``Clustering social
  networks,'' in \emph{Algorithms and Models for the Web-graph}.\hskip 1em plus
  0.5em minus 0.4em\relax Springer, 2007, pp. 56--67.

\bibitem{gupta2011evolutionary}
M.~Gupta, C.~C. Aggarwal, J.~Han, and Y.~Sun, ``Evolutionary clustering and
  analysis of bibliographic networks,'' in \emph{International Conference on
  Advances in Social Networks Analysis and Mining (ASONAM), 2011}.\hskip 1em
  plus 0.5em minus 0.4em\relax IEEE, 2011, pp. 63--70.

\bibitem{bilgin2006dynamic}
C.~C. Bilgin and B.~Yener, ``Dynamic network evolution: Models, clustering,
  anomaly detection,'' \emph{IEEE Networks}, 2006.

\bibitem{brandes:2001}
U.~Brandes, ``A faster algorithm for betweenness centrality.'' \emph{Journal of
  Mathematical Sociology}, vol.~25, pp. 163--177, 2001.

\bibitem{gould:1967}
G.~P. R., ``On the geographical interpretation of eigenvalues,''
  \emph{Transactions of the Institute of British Geographers}, no.~42, pp.
  53--86, Dec. 1967.

\bibitem{beauchamp:1965}
M.~A. Beauchamp, ``An improved index of centrality.'' \emph{Behavioral
  Science}, vol.~1-, pp. 161--163, 1965.

\bibitem{coremen:2001}
T.~H. Coremen, C.~E. Leiserson, R.~L. Rivest, and C.~Stein, \emph{Introduction
  to Algorithms}.\hskip 1em plus 0.5em minus 0.4em\relax The MIT Press, 2001.

\bibitem{freeman1:1977}
L.~C. Freeman, ``A set of measures of centrality based on betweenness,''
  \emph{Sociometry}, 1977.

\bibitem{wen:2013}
S.~Wen, W.~Zhou, J.~Zhang, Y.~Xiang, W.~Zhou, and W.~Jia, ``Modeling
  propagation dynamics of social network worms,'' \emph{Parallel and
  Distributed Systems, IEEE Transactions on}, vol.~24, no.~8, pp. 1633--1643,
  Aug 2013.

\end{thebibliography}

\end{document}